\documentclass{article}%
\usepackage{amsfonts}
\usepackage{amsmath}
\usepackage{amssymb}
\usepackage{graphicx}%
\providecommand{\U}[1]{\protect\rule{.1in}{.1in}}

\begin{document}

\title{\textbf{Preferences in Quantum Games}}
\author{\textit{Simon J.D. Phoenix \& Faisal Shah Khan}\\Khalifa University, PO Box 127788, Abu Dhabi, UAE }
\maketitle

\begin{abstract}
A quantum game can be viewed as a state preparation in which the final output
state results from the competing preferences of the players over the set of
possible output states that can be produced. It is therefore possible to view
state preparation in general as being the output of some appropriately chosen
(notional) quantum game. This reverse engineering approach in which we seek to
construct a suitable notional game that produces some desired output state as
its equilibrium state may lead to different methodologies and insights. With
this goal in mind we examine the notion of preference in quantum games since
if we are interested in the production of a particular equilibrium output
state, it is the competing preferences of the players that determine this
equilibrium state. We show that preferences on output states can be viewed in
certain cases as being induced by measurement with an appropriate set of
numerical weightings, or payoffs, attached to the results of that measurement.
In particular we show that a distance-based preference measure on the output
states is equivalent to a having a strictly-competitive set of payoffs on the
results of some measurement.

\end{abstract}

\section{Introduction}

Whilst there are important issues to resolve in our full understanding of the
relationship between correlation, entanglement and non-classical properties,
it is clear that quantum mechanics admits possibilities for behaviours that
cannot be achieved by any purely classical system. The introduction of quantum
mechanical ideas in game theory is one example where entanglement offers the
potential to achieve game results that cannot be obtained when playing games
with classical objects obeying the laws of classical physics [1,2].

In our previous work we have argued for the importance of the notion of
\textit{playable} games; that is, games that can actually be played, or
implemented, with objects that have a physical reality and obey physical laws
[3-5]. This perspective allows a general framework for the description of a
game (either quantum or classical) to be developed in which the comparisons
between classical and quantum behaviours can more easily be drawn. In this
perspective a game, either quantum or classical, can be seen as a
\textit{state preparation} where the state that is actually prepared is a
function of the initial state, the actions available to the players and their
competing preferences over the possible states that can be produced.

If games are \textit{state preparations}, then might not the converse be true?
Can we view state preparations as the output of certain quantum games? If this
is possible then we can view state preparations from a game-theoretic
perspective. Instead of asking the question whether quantum games give us
results unobtainable in classical games, the shift in view to state
preparation asks the question whether we can re-interpret quantum behaviours
in a game-theoretic way. In our previous work we have expressed this notion as
that of `gaming the quantum' rather than the usual approach in quantum games
of `quantizing the game' [3]. In this way we ask whether the tools of game
theory can be applied to quantum mechanics, just as they are applied to fields
like economics or biology, in order to give us different insights or different
computational methodologies.

With this goal in mind therefore, it is important to construct the right
conceptual framework for working within this perspective. If we're interested
in state preparations then the states that are prepared result from the
competing actions of the players. The players must have some reason, some
preference, for the preparation of one state over another. Therefore the
appropriate language is that of \textit{preference}. The players have some
preference for the states they wish to be output from a game and it is the
competing preferences of the players that result in a particular output state.

Accordingly, in this paper we wish to explore the notion of preference in
quantum games. It is common in game theory to associate numerical weightings,
which for clarity we term payoffs, with measurement results\footnote{Here we
make explicit that even in a classical game the output state produced by the
players in a playable game has to be measured. Of course in most classical
games this can be safely ignored since measurement does not disturb the output
(pre-measurement) state produced by the players.} to express a preference;
each player attempting to act in such a way as to maximise the numerical value
of his or her payoff. Thus in classical Prisoner's Dilemma we might envision
coins initially prepared in the state $\left(  H,H\right)  $ with the players
each given one coin that they can flip or leave alone. The players will
produce one of 4 possible output states of the coins which can then be
determined (measured) and the measurement result mapped to some specific
payoff. For example, if the players produce the output state $\left(
H,H\right)  $ then the payoff for the players is commonly expressed as the
tuple $\left(  3,3\right)  $ and the output state $\left(  H,T\right)  $ is
associated with the payoff tuple $\left(  0,5\right)  $. In classical games
the act of measurement is, in general, taken to be a passive action which does
not affect the output state produced by the players. The resulting matrix of
payoff tuples for the possible actions of the players then determines the
particular actions that will be chosen, assuming rational players.

In quantum mechanics, however, the act of measurement is anything but a
passive process. If we consider an ideal von Neumann measurement then, unless
the output states produced by the players are eigenstates of the measurement
operator, the measurement will project the output state of the players onto a
new state according to some probability distribution. The quantum measurement
thus furnishes an \textit{expected} payoff for the players. As we have shown
[4], if the players act in such a way as to maximize their expected payoffs,
this can lead to a transformation of a game in the sense that although the
players initially have some preference over the measurement eigenstates
(expressed by some numerical weighting) these lead to an \textit{induced}
preference over the output states as given by the ordering of the expected
payoffs. The initial weightings associated with the \textit{measurement}
results may be consistent with a game of Prisoner's Dilemma, say, but the
players make their choice according to an analysis of the \textit{expected
payoffs} and may end up playing a game with the preferences of Chicken rather
than Prisoner's Dilemma.

This feature of game transformation, as decribed above, is not unique to
quantum mechanics and can occur in strictly classical games in which there is
a distribution over measurement results. Such a distribution can occur, for
example, if the choices of the players are communicated over a noisy channel.
The important feature is that the initial preferences over the measurement
results, expressed by numerical weightings, \textit{induce} a preference on
the pre-measurement output states. It is these \textit{induced} preferences
that determine the game the players \textit{actually} play and not the initial
preferences on the measurement eigenstates.

In this paper we explore a description of quantum games in terms of
preferences over the pre-measurement output states. We show that a game in
which there are strictly competitive preferences defined on the measurement
eigenstates is equivalent to a game in which the preferences over the
pre-measurement output states are defined according to a distance from some
most-preferred state. As we have previously shown, preferences on output
states based on distance have a nice geometric interpretation [3]. Our result
here shows that any quantum game with a strictly-competitive weighting on the
measurement eigenstates can be interpreted using this geometrical treatment on
the pre-measurement output states.

\section{General Formulation of a Discrete Quantum Game}

We define a Discrete Quantum Game (DQG) as one in which the players have a
finite set of actions from which to choose; the actions being unitary
transformations of the input state. There will, therefore, be a finite set of
possible output states produced by the players. Of course, in many quantum
game formulations the players can access a continuum of possibilities (such as
a general rotation of a qubit). For convenience we restrict our analysis to
the case of a DQG, but the main results carry over to the continuum case in a
straightforward manner.

The basic elements are the following:

\begin{itemize}
\item There is some initial state $\left\vert \psi_{in}\right\rangle $

\item The players have some finite set of available operations with which they
can modify the input state. We denote the operations available to player 1 by
$\left\{  \hat{\alpha}_{1},\ldots,\hat{\alpha}_{m}\right\}  $ and those
available to player 2 by $\left\{  \hat{\beta}_{1},\ldots,\hat{\beta}%
_{n}\right\}  $

\item The set of possible output states that can be produced is denoted by
$\Psi_{out}$ and there are up to a total of $mn$ possible distinct output
states that can be produced. In other words $\left\vert \Psi_{out}\right\vert
\leq mn$

\item $\Psi_{out}\subset\Psi$ where $\Psi$ is the set of all possible states
of the physical system on which the players operate
\end{itemize}

The final element required to be able to describe this as a game is that the
players have some preference over the output states that are produced. If the
set of output states is written as%
\[
\Psi_{out}=\left\{  \left\vert \psi_{1}\right\rangle ,\left\vert \psi
_{2}\right\rangle ,\ldots,\left\vert \psi_{mn}\right\rangle \right\}
\]
then the preferences of each player are equivalent to different orderings of
this set. The players thus choose an operation from their available set in
order to produce their most preferred output state \textit{given that their
opponent is doing the same}. It is this `push-pull' on the possible output
states that leads to the production of an equilibrium state, where such an
equilibrium exists. The general formulation of a DQG is shown in figure 1

\begin{quote}%
\[%
{\includegraphics[
natheight=7.499600in,
natwidth=9.999800in,
height=3.8329in,
width=5.1041in
]%
{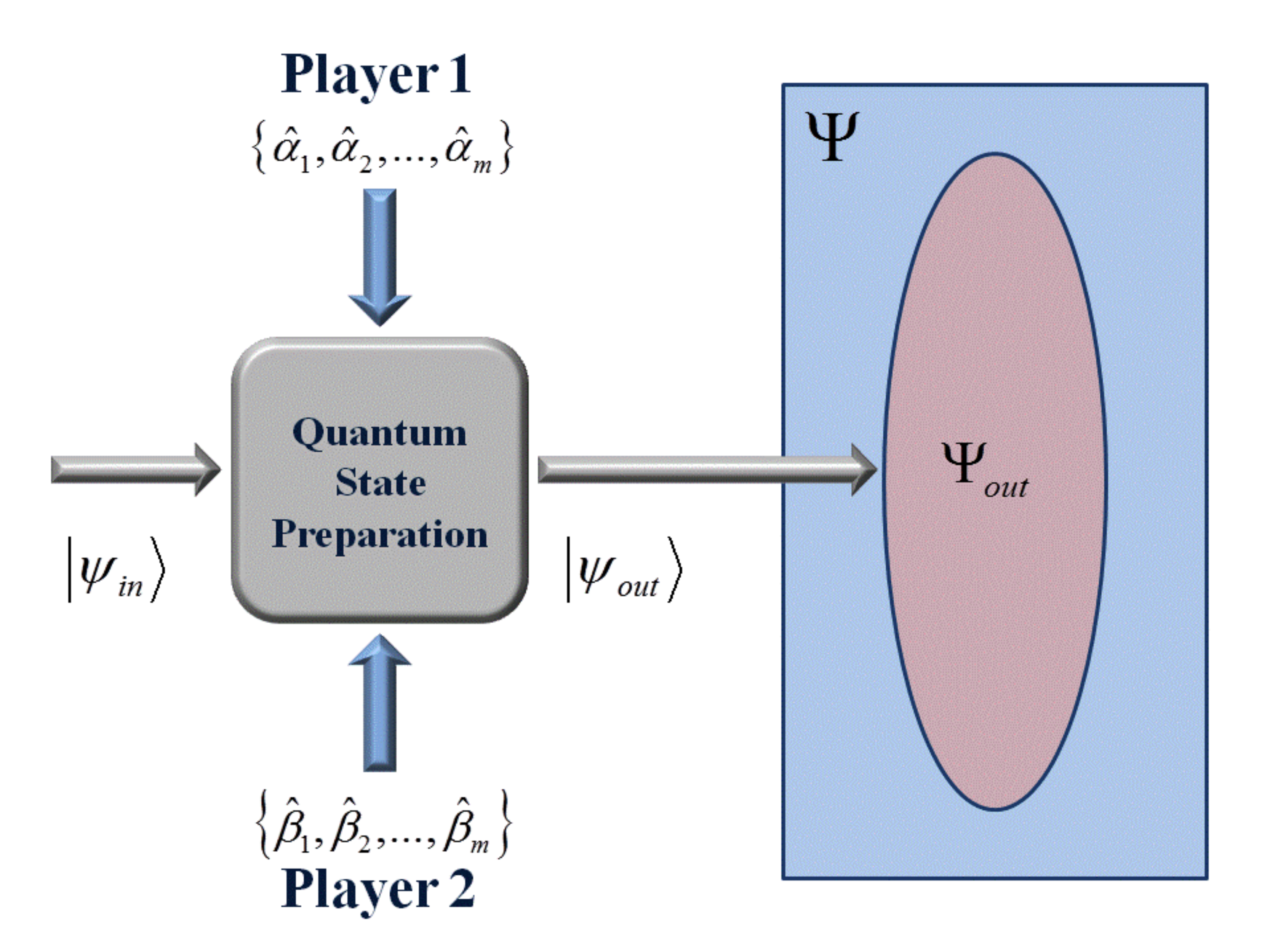}%
}
\]
\textbf{Figure 1 }: \textit{the general description of a discrete quantum game
(DQG). The players each have a finite set of available unitary operations
which act on some initial state. The players have competing preferences over
the set of possible output states }$\Psi_{out}$ \textit{that can be prepared
\ The actual state }$\left\vert \psi_{in}\right\rangle \in\Psi_{out}$
\textit{that is prepared depends on the theoretical analysis the players
undertake of each others' choices based upon their knowledge of the available
operations and preferences}.
\end{quote}

\subsection{Quantum Formulation of Standard Prisoner's Dilemma}

As an example of the application of this formalism we consider the
implementation of the standard version of Prisoner's Dilemma (PD) using 2
spin-1/2 particles. The initial state is expressed in the computational basis
(taken to be eigenstates of the spin-$z$ operator) as $\left\vert
00\right\rangle $. The players each have a particle on which to act and the
available operations are $\left\{  \hat{\alpha}_{1}=\hat{I}_{1},\hat{\alpha
}_{2}=\hat{F}_{1}\right\}  $ and $\left\{  \hat{\beta}_{1}=\hat{I}_{2}%
,\hat{\beta}_{2}=\hat{F}_{2}\right\}  $ for players $A$ and $B$, respectively,
and $\hat{I}$ and $\hat{F}$ are the identity and spin-flip operators acting on
particles 1 and 2. There are thus 4 possible output states that can be
produced which can be labelled as%
\begin{align}
\left\vert \psi_{1}\right\rangle  &  =\hat{\alpha}_{1}\hat{\beta}%
_{1}\left\vert \psi_{in}\right\rangle =\left\vert 00\right\rangle \nonumber\\
\left\vert \psi_{2}\right\rangle  &  =\hat{\alpha}_{1}\hat{\beta}%
_{2}\left\vert \psi_{in}\right\rangle =\left\vert 01\right\rangle \nonumber\\
\left\vert \psi_{3}\right\rangle  &  =\hat{\alpha}_{2}\hat{\beta}%
_{1}\left\vert \psi_{in}\right\rangle =\left\vert 10\right\rangle \nonumber\\
\left\vert \psi_{4}\right\rangle  &  =\hat{\alpha}_{2}\hat{\beta}%
_{2}\left\vert \psi_{in}\right\rangle =\left\vert 11\right\rangle
\end{align}
In PD the preferences over these output states can be expressed for players
$A$ and $B$ as%
\begin{align}
A  &  :~~~\left\vert \psi_{3}\right\rangle \succ\left\vert \psi_{1}%
\right\rangle \succ\left\vert \psi_{4}\right\rangle \succ\left\vert \psi
_{2}\right\rangle \nonumber\\
B  &  :~~~\left\vert \psi_{2}\right\rangle \succ\left\vert \psi_{1}%
\right\rangle \succ\left\vert \psi_{4}\right\rangle \succ\left\vert \psi
_{3}\right\rangle
\end{align}
It is common to express the preferences in terms of some numerical weighting
over the measurement results with respect to some measurement operator. If a
measurement of spin-$z$ is made on both particles the set of eigenstates of
the measurement operator is equal to the set of possible output states that
the players produce. The usual weightings for PD, given by a mapping of a
numerical value to the measurement states, are%
\begin{align}
\left\vert 00\right\rangle  &  \longrightarrow\left(  3,3\right) \nonumber\\
\left\vert 01\right\rangle  &  \longrightarrow\left(  0,5\right) \nonumber\\
\left\vert 10\right\rangle  &  \longrightarrow\left(  5,0\right) \nonumber\\
\left\vert 11\right\rangle  &  \longrightarrow\left(  1,1\right)
\end{align}
There is nothing particularly quantum mechanical about this form of PD.
Indeed, we could view this quantum formulation of the game as an expensive
implementation of the classical version.

In this formulation of the game the possible output states that can be
produced by the players are the \textit{eigenstates} of the measurement
operator. However, we are not restricted to consideration of this particular
measurement. The possible output states may not be eigenstates of the
subsequent measurement operator\footnote{Consider, for example, a measurement
of spin at some angle $\theta$ with respect to $z$. The set of measurement
results will be the eigenstates $\left\{  \left\vert 00\right\rangle _{\theta
},\left\vert 01\right\rangle _{\theta},\left\vert 10\right\rangle _{\theta
},\left\vert 11\right\rangle _{\theta}\right\}  $ the particular state
obtained upon measurement being determined by a probability distribution
dependent upon $\theta$. If we establish some preference over these
measurement results via a mapping to a tuple of outcomes for each eigenstate,
then the expected outcome tuples for the players will induce a set of
preferences on the output states. In the extreme case, $\theta=\pi/2$, each
possible output state is equally preferred via this induced preference.}. The
measurement will then yield a measurement result with a particular
probability. In order to translate this to a preference on the output states
we must consider some measure such as the \textit{expected} payoffs for the
players for each possible output state.

In general, therefore, a set of preferences that is \textit{induced} on the
output states via application of the mapping of numerical weights to the
measurement states and consideration of the expected outcome \textit{need not
be equal} to the initial preferences over the measurement results. This is
obvious from the DQG formalism since if we denote the set of measurement
eigenstates as $\Psi_{meas}$ then the cardinalities of the sets $\Psi_{meas}$
and $\Psi_{out}$ are only equal in special cases.

\subsection{Games Using the EWL Protocol}

In their ground-breaking paper on quantum games Eisert, Wilkens and Lewenstein
(EWL) proposed the following structure for a game played with 2 qubits
represented by a pair of spin-1/2 particles [2]. The qubits are prepared in
some initial state (for convenience often taken to be the ground state
$\left\vert 00\right\rangle $ with respect to some spin direction). An
entanglement operator $\hat{E}$ is applied to the qubits and each player
receives on of the particles. The players can then apply some local unitary
rotation to their particle which we denote by $\hat{U}_{A\left(  B\right)  }$.
A disentangling operator $\hat{E}^{-1}$ is then applied to the qubits and a
measurement of spin in some given direction is made on each particle. A set of
weightings on the output states is given which reflect the classical game
which the quantum version generalizes.

This protocol has become the \textit{de facto} standard in treatments of
quantum games (see, for example, the brief selection [6-12] which illustrate
some interesting and important results obtained using this protocol). The
entangling/disentangling step is to exploit the resource of entanglement
whilst still allowing the players to perform local operations on their
respective particle. It is a clever protocol which reduces to the classical
game (described by the measurement weightings) in the appropriate limit [13].
The perception that the players only act on their own particle is, however,
something of an illusion. The DQG formalism for playable games shows that in
reality the players are acting on the input state, which in the case of the
EWL protocol is a maximally entangled state. The EWL protocol is equivalent to
a game in which the players are given the set of available operations
$\left\{  \hat{E}^{-1}\hat{U}_{A}\hat{E}\right\}  $ and $\left\{  \hat{E}%
^{-1}\hat{U}_{B}\hat{E}\right\}  $, for players \textit{A} and \textit{B},
respectively. In effect in games of this structure the players act on
\textit{both} particles.

It is clear in the EWL protocol that we have only 4 measurement eigenstates
but a continuum of available operations for the players, and therefore a
continuum of possible output states. It is important that the available
operations available to the players do not provide a map of the initial state
onto the space of all possible 2 qubit states, $\Psi$, otherwise equilibrium
cannot be attained. The operator sets $\left\{  \hat{E}^{-1}\hat{U}_{A}\hat
{E}\right\}  $ and $\left\{  \hat{E}^{-1}\hat{U}_{B}\hat{E}\right\}  $ only
produce a subset of the set of possible 2 qubit states. Indeed, in the case of
the EWL protocol, the available operations cannot access the set of maximally
entangled states of 2 qubits and so $\Psi_{out}\subset\Psi$.

The weightings on the measurement eigenstates \textit{induce} a set of
preferences on this subset via the expected payoffs for the players. It is
then clear why EWL results in an enhanced equilibrium payoff for the players
for Prisoner's Dilemma; we are no longer playing standard PD but a different
game\footnote{It is appropriate to view quantum PD as an extension of standard
PD implemented with quantum objects. If we then wish to compare quantum and
classical features we must consider comparison with possible \textit{classical
extensions} of PD.}. It is possible to find a classical extension of PD with
correlated classical noise which gives the same enhancement of the equilibrium
payoff [4].

\subsection{Non-Commuting Games}

The DQG treatment of the EWL protocol highlights another interesting feature
of quantum games. It is possible to consider games in which the operations
available to the players do not commute. In terms of games in which we have 2
qubits we then envision a game in which the players have the capability of
affecting each other's particle. So we can define a non-commuting game as a
game in which $\left[  \hat{\alpha}_{i},\hat{\beta}_{j}\right]  \neq0$ for at
least one choice of $i$ and $j$.

In the DQG formalism we can see that the players act on some initial quantum
state $\left\vert \psi_{in}\right\rangle $ and so, in general, the players act
on a single entity in quantum mechanical terms. It is only in a special class
of quantum games where the Hilbert space can be split into a physically
meaningful tensor product $H_{A}\otimes H_{B}$ can we envision the possibility
of the players acting on `independent' physical entities. As we have seen, in
the EWL protocol the players only have the illusion of acting on separate
particles; the entanglement operation giving the necessary connection so that
game transformation (and enhancement of equilibrium payoff) can be achieved.
The standard EWL protocol is a commuting game; the players only have access to
local operations after the initial entanglement and $\left[  \hat{E}^{-1}%
\hat{U}_{A}\hat{E}~,~\hat{E}^{-1}\hat{U}_{B}\hat{E}\right]  =0$ for any choice
of $\hat{U}_{A}$ and $\hat{U}_{B}$.

Thus, in general, in non-commuting games, the order of play matters. It is
easy to construct simple game examples in which swtiching the order of play
changes the game that is played. This is obvious from the formalism of a DQG
since the set of output states where $B$ plays first is $\Psi_{out}%
^{AB}=\left\{  \hat{\alpha}_{i}\hat{\beta}_{j}\left\vert \psi_{in}%
\right\rangle \right\}  $ and the set of output states where $A$ plays first
is $\Psi_{out}^{BA}=\left\{  \hat{\beta}_{j}\hat{\alpha}_{i}\left\vert
\psi_{in}\right\rangle \right\}  $ so that for non-commuting games $\Psi
_{out}^{AB}\neq\Psi_{out}^{BA}$.

\subsection{Time Evolution as a Quantum Game}

As another illustration of the approach based on preferences and the DQG
perspective consider the following simple and somewhat formal game in which
the operator sets available to the players are%
\begin{align}
&  \left\{  \hat{I},\exp\left(  -it\left\{  \frac{\hat{H}+\Delta\hat{H}}%
{2}\right\}  \right)  \right\}  _{A}\nonumber\\
& \nonumber\\
&  \left\{  \hat{I},\exp\left(  -it\left\{  \frac{\hat{H}-\Delta\hat{H}}%
{2}\right\}  \right)  \right\}  _{B}%
\end{align}
where $\hat{I}$ is the identity, $\hat{H}$ is a time-independent Hamiltonian
and we suppose that $\left[  \hat{H},\Delta\hat{H}\right]  =0$. If our input
state to the game is $\left\vert \psi\left(  0\right)  \right\rangle $ then
the 4 possible output states are $\left\{  \left\vert \psi\left(  0\right)
\right\rangle ,\left\vert \psi_{+}\left(  t\right)  \right\rangle ,\left\vert
\psi_{-}\left(  t\right)  \right\rangle ,\left\vert \psi\left(  t\right)
\right\rangle \right\}  $ where $\left\vert \psi_{\pm}\left(  t\right)
\right\rangle =\exp\left(  -it\left\{  \frac{\hat{H}\pm\Delta\hat{H}}%
{2}\right\}  \right)  \left\vert \psi\left(  0\right)  \right\rangle $ and
$\left\vert \psi\left(  t\right)  \right\rangle =\exp\left(  -i\hat
{H}t\right)  \left\vert \psi\left(  0\right)  \right\rangle $. If the players
have preferences on the output states given by
\begin{align}
A &  :~~~\left\vert \psi_{+}\right\rangle \succ\left\vert \psi\left(
t\right)  \right\rangle \succ\left\vert \psi\left(  0\right)  \right\rangle
\succ\left\vert \psi_{-}\right\rangle \nonumber\\
B &  :~~~\left\vert \psi_{-}\right\rangle \succ\left\vert \psi\left(
0\right)  \right\rangle \succ\left\vert \psi\left(  t\right)  \right\rangle
\succ\left\vert \psi_{+}\right\rangle
\end{align}
then the Nash equilibrium output of this game is $\left\vert \psi\left(
t\right)  \right\rangle $ which is simply the time-evolution of the initial
state under the Hamiltonian $\hat{H}$. This is admittedly a rather contrived
and artificial example, but it does illustrate the possibility of interpreting
the evolution of a quantum state as the output of some 2-player game. Of
course, it is clear that there will be an infinite number of possible games we
can construct that yield the desired time-evolved state as the equilibrium
output. If the available operations, and preferences, are suitably chosen then
we may obtain an insight into quantum evolution as a `push-pull' between
competing physical processes. Rather speculatively, then, we might ask whether
it is possible to view the path integral approach to quantum mechanics as a
competition between different paths giving an interpretation of the least
action principle as arising from a game-theoretic description. \ 

\bigskip\bigskip

If the preferences over the output states determine the game the players
actually play, then in a certain sense it is these preferences that have a
more fundamental character than preferences expressed as numerical weightings
over some measurement results with respect to some given operator; different
measurement operators may induce different preferences on the possible output
states that can be produced. By shifting the perspective to preferences over
the output states, rather than preferences over some specific set of
measurement results we obtain a more general description of a quantum game.
The players are not restricted to obtaining their preferences on the possible
output states via consideration of a specific measurement and the induced
preferences over those output states, but can use whatever rule they see fit
to generate their preference. The only thing that is important, as far as the
analysis of the state preparation as a game is concerned, is that the players
have \textit{some} preference over the output states. \textit{From where they
get their preference over the output states is largely immaterial}.

\section{Operations and Preferences}

In the perspective we have outlined above a game is completely specified by

\begin{itemize}
\item the initial state $\left\vert \psi_{in}\right\rangle $

\item the available operations specified by the sets $\left\{  \hat{\alpha
}_{1},\ldots,\hat{\alpha}_{m}\right\}  $ and $\left\{  \hat{\beta}_{1}%
,\ldots,\hat{\beta}_{n}\right\}  $

\item an ordering, for each player, on the possible output states $\left\vert
\psi_{ij}\right\rangle =\hat{\alpha}_{i}\hat{\beta}_{j}\left\vert \psi
_{in}\right\rangle \in\Psi_{out}$ which expresses their preferences on this
set. (Equivalently, for a given initial state, we can view the preferences as
orderings on the set of operations with elements $\hat{\alpha}_{i}\hat{\beta
}_{j}$)
\end{itemize}

A game can then be viewed as an algorithmic procedure which takes as inputs
the initial state, the sets of operations, and the preferences over the states
$\left\vert \psi_{ij}\right\rangle =\hat{\alpha}_{i}\hat{\beta}_{j}\left\vert
\psi_{in}\right\rangle $ and outputs a tuple $\left\{  \hat{\alpha}_{p}%
,\hat{\beta}_{s}\right\}  $. If this algorithmic procedure does not terminate
then there is no equilibrium choice for the players. It is important to note
that the preferences on the \textit{output states} alone are not sufficient to
determine the game that is played. This is nicely illustrated with the
following example.

Consider 2 players playing a game which has a physical implementation using 2
spin-1/2 particles. The initial state is expressed in the computational basis
(taken to be eigenstates of the spin-$z$ operator) as $\left\vert
00\right\rangle $. The operations available to the players are $\left\{
\hat{\alpha}_{1},\hat{\alpha}_{2}\right\}  $ and $\left\{  \hat{\beta}%
_{1},\hat{\beta}_{2}\right\}  $ for players $A$ and $B$, respectively. There
are thus 4 possible output states that can be produced. Let us consider
operations such that we have the set of output states%
\begin{align}
\left\vert \psi_{1}\right\rangle  &  =\hat{\alpha}_{1}\hat{\beta}%
_{1}\left\vert \psi_{in}\right\rangle =\left\vert 00\right\rangle \nonumber\\
\left\vert \psi_{2}\right\rangle  &  =\hat{\alpha}_{1}\hat{\beta}%
_{2}\left\vert \psi_{in}\right\rangle =\left\vert 01\right\rangle \nonumber\\
\left\vert \psi_{3}\right\rangle  &  =\hat{\alpha}_{2}\hat{\beta}%
_{1}\left\vert \psi_{in}\right\rangle =\left\vert 10\right\rangle \nonumber\\
\left\vert \psi_{4}\right\rangle  &  =\hat{\alpha}_{2}\hat{\beta}%
_{2}\left\vert \psi_{in}\right\rangle =\left\vert 11\right\rangle
\end{align}
If we denote the operation of flipping the spin of particle 1 by $\hat{F}_{1}
$ and the same operation for particle 2 as $\hat{F}_{2}$, then \textit{one
way} to achieve these possible output states is to give the players the
choices $\left\{  \hat{\alpha}_{1}=\hat{I},\hat{\alpha}_{2}=\hat{F}%
_{1}\right\}  $ and $\left\{  \hat{\beta}_{1}=\hat{I},\hat{\beta}_{2}=\hat
{F}_{2}\right\}  $ where $\hat{I}$ is the identity operation. If the players
have the preferences defined by the orderings of $\Psi_{out}$%

\begin{align}
A &  :~~~\left\vert \psi_{1}\right\rangle \succ\left\vert \psi_{2}%
\right\rangle \succ\left\vert \psi_{3}\right\rangle \succ\left\vert \psi
_{4}\right\rangle \nonumber\\
B &  :~~~\left\vert \psi_{4}\right\rangle \succ\left\vert \psi_{3}%
\right\rangle \succ\left\vert \psi_{2}\right\rangle \succ\left\vert \psi
_{1}\right\rangle
\end{align}
then it is evident that the equilibrium state $\left\vert \psi_{2}%
\right\rangle =\hat{\alpha}_{1}\hat{\beta}_{2}\left\vert \psi_{in}%
\right\rangle =\left\vert 01\right\rangle $ is produced. However, if we retain
these same preferences over the output states but consider operators such that%
\begin{align}
\left\vert \psi_{1}\right\rangle  &  =\hat{\alpha}_{1}\hat{\beta}%
_{1}\left\vert \psi_{in}\right\rangle =\left\vert 00\right\rangle \nonumber\\
\left\vert \psi_{2}\right\rangle  &  =\hat{\alpha}_{2}\hat{\beta}%
_{1}\left\vert \psi_{in}\right\rangle =\left\vert 01\right\rangle \nonumber\\
\left\vert \psi_{3}\right\rangle  &  =\hat{\alpha}_{1}\hat{\beta}%
_{2}\left\vert \psi_{in}\right\rangle =\left\vert 10\right\rangle \nonumber\\
\left\vert \psi_{4}\right\rangle  &  =\hat{\alpha}_{2}\hat{\beta}%
_{2}\left\vert \psi_{in}\right\rangle =\left\vert 11\right\rangle
\end{align}
then the equilibrium state $\left\vert \psi_{3}\right\rangle =\hat{\alpha}%
_{1}\hat{\beta}_{2}\left\vert \psi_{in}\right\rangle =\left\vert
10\right\rangle $ is produced by the game. We can achieve this game by giving
the players the sets of operator choices%
\begin{align}
A &  :~~~\left\{  \hat{\alpha},\hat{F}_{2}\hat{\alpha}\right\}  \nonumber\\
B &  :~~~\left\{  \hat{\alpha}^{-1},\hat{\alpha}^{-1}\hat{F}_{1}\right\}
\end{align}
where $\hat{\alpha}$ is any arbitrary unitary transformation that acts on the
particles. If we think in terms of an implementation in which each player is
given a ball which can be switched from a red or blue colour, with the initial
state being that both balls are red, then this latter game is equivalent to
giving the players the ability to play with each other's balls. It is easy to
see from these particular physical instantiations of the games that they are,
in fact, different games. This is perhaps more clearly seen by examining the
preferences over the available operations. In the former game we have the
preferences for player \textit{A} over the available operations given by
\begin{equation}
A:~~~\hat{\alpha}_{1}\hat{\beta}_{1}\succ\hat{\alpha}_{1}\hat{\beta}_{2}%
\succ\hat{\alpha}_{2}\hat{\beta}_{1}\succ\hat{\alpha}_{2}\hat{\beta}_{2}%
\end{equation}
whereas in the latter game we have the preferences%
\begin{equation}
A:~~~\hat{\alpha}_{1}\hat{\beta}_{1}\succ\hat{\alpha}_{2}\hat{\beta}_{1}%
\succ\hat{\alpha}_{1}\hat{\beta}_{2}\succ\hat{\alpha}_{2}\hat{\beta}_{2}%
\end{equation}
Thus whilst we may have the same preferences over the output states in both
games we have different preferences over the set of available operations. It
is clear from this simple example that there will be many ways to assign
operations to the players that lead to the same set of output states. The same
set of preferences on output states can, therefore, represent different games.
With the proviso that it is in fact a more fundamental perspective to consider
preferences over the set of \textit{operations} we shall continue to examine
the preferences over the \textit{output states} in the subsequent sections,
assuming that the available set of operations has been fixed.

\section{Preferences in Discrete Quantum Games}

In much of the previous work on quantum game theory the preferences over the
output states have largely been derived by reference to some subsequent
measurement that is performed on the output states. The numerical weightings
assigned to the possible measurement results then \textit{induce} preferences
on the pre-measurement output states via some measure such as the expected
outcome for the players. All that is actually required for the players to
determine their choice of strategy is that they have some preference on the
possible pre-measurement output states that can be produced. The players
(assumed rational in their choice of action) must have some foundation for the
\textit{theoretical} analysis they undertake in order to decide their choice
of operation, \textit{before any measurement or operation is performed}, and
that foundation is provided by the preferences on the possible output states.

Of course each player is free to choose a preference over the output states in
any arbitrary way. To illustrate this in a particularly gruesome fashion we
could imagine that one player determines his preferences at random; the random
numbers being generated by removing the legs from one side of a spider and
noting the time it takes for the hapless spider to turn 3
circles\footnote{Players are only assumed rational in their analysis of the
preferences that lead to their choice of action. The players are not required
to have a rational basis for their choice of preference.}. We are not
interested in such whimsical or bizarre methods for determining a preference
but in those preferences that can be generated by a well-defined prescription
that is related to quantities of physical interest. Examples of such
prescriptions may include

\begin{itemize}
\item A known measurement is to be performed. Numerical weightings are
assigned to the possible measurement results and the players derive their
preferences by ranking the output states in order of decreasing expected
outcomes. This analysis is, obviously, performed pre-measurement. This is a
common way to assign preferences on the output states in a quantum game; the
preferences being induced by the assignment of weights on the post-measurement
states.and the calculation of an expected outcome.

\item A preference may be defined by ranking the output states in terms of
their distance from some state $\left\vert \xi\right\rangle \in\Psi$. This
chosen state $\left\vert \xi\right\rangle $ may, or may not be, a member of
the set $\Psi_{out}$. It is necessary that the players choose to assign their
preferences with respect to different states otherwise both players generate
identical preferences over the output states and we do not have a game. Thus
player $A$ may be interested in minimizing the distance of the output states
from some arbitrary state $\left\vert \xi_{A}\right\rangle $ whereas player
$B$ may minimize the distance with respect to some state $\left\vert \xi
_{B}\right\rangle $

\item Player $A$ may assign his preference on the output states by requiring
that the variance of the output states with respect to some measurement
$\hat{M}_{A}$ is minimized. Player $B$ may seek to produce an output state
that minimizes the variance with respect to some measurement $\hat{M}_{B}$.
Again, it is necessary that the players seek to minimize the variance with
respect to different operators otherwise they generate an identical ordering
(preference) on the output states.

\item The players may wish to choose an output state that maximizes the
Shannon information with respect to some operator. Again the operator choices
made by the players, for the purpose of defining their preference, should not
be the same.
\end{itemize}

These few examples are clearly not an exhaustive list of all the interesting,
and physically relevant, ways in which the players can generate their
preferences on the output states. As these examples illustrate, it is
necessary that the players generate a \textit{different} ordering (preference)
on the set of possible pre-measurement output states.

As a possible speculative example of this approach let us consider a quantum
computation that produces some desired output state $\left\vert \lambda
\right\rangle $. If we identify this as the Nash equilibrium state of some
game between 2 players then the challenge is to find a game such that it
produces the desired computation as an equilibrium output for any choice of
input. The preferences over the output states may then be translated to
preferences over some measurement on the output states. If we can find such a
game, and such a measurement, then we have a model of a quantum computation as
a 2-player game. It may be technically easier to implement this game than to
build the various quantum logic gates required. Whilst it appears possible to
theoretically construct such games and measurements for very simple quantum
computations it seems to be highly non-trivial to demonstrate this in general.

\subsection{Preferences Induced By Measurement}

We now consider a situation where the players determine their preferences over
the output states by analysis of the results of a subsequent measurement. As
above we assume some known input state $\left\vert \psi_{in}\right\rangle $
upon which the players can act with a discrete set of operations specified by
the sets $\left\{  \hat{\alpha}_{1},\ldots,\hat{\alpha}_{m}\right\}  $ and
$\left\{  \hat{\beta}_{1},\ldots,\hat{\beta}_{n}\right\}  $. There are thus up
to $mn$ possible pre-measurement output states that can be produced given by%
\[
\left\vert \psi_{ij}\right\rangle =\hat{\alpha}_{i}\hat{\beta}_{j}\left\vert
\psi_{in}\right\rangle
\]

It is often assumed that the Hilbert space of the initial input system is the
product space $H=H_{A}\otimes H_{B}$ in order to allow each player to act on a
separate physical entity (the typical scenario being the space of 2 spin-1/2
particles representing 2 qubits). This division is not strictly necessary and
if we allow the possibility of entanglement (either through specification of
an initial state or actions allowed by the players) then, strictly speaking,
we must consider the 2 particles as a single entity defined by a single state
as we have described above.

\begin{itemize}
\item The subsequent measurement on the output states is described by the
operator%
\begin{equation}
\hat{M}=%
{\displaystyle\sum\limits_{k}}
\varphi_{k}\left\vert \varphi_{k}\right\rangle \left\langle \varphi
_{k}\right\vert
\end{equation}
where $\hat{M}\left\vert \varphi_{k}\right\rangle =\varphi_{k}\left\vert
\varphi_{k}\right\rangle $ and we assume an ideal measurement. The set of
measurement eigenstates will be denoted by $\Psi_{M}$.

\item The measurement of $\hat{M}$, made on the output state $\left\vert
\psi_{ij}\right\rangle =\hat{\alpha}_{i}\hat{\beta}_{j}\left\vert \psi
_{in}\right\rangle $, results in the eigenstate $\left\vert \varphi
_{k}\right\rangle $\ with probability $\left\vert \left\langle \psi
_{ij}\right.  \left\vert \varphi_{k}\right\rangle \right\vert ^{2}$

\item To each eigenstate $\left\vert \varphi_{k}\right\rangle $\ a tuple is
assigned, denoted by $\left(  \omega_{k}^{A},\omega_{k}^{B}\right)  $ which
gives the payoff for the players, for that particular measurement result

\item The expected payoffs for the players, for each possible pre-measurement
output state, are given by%
\begin{align}
\left\langle \hat{O}_{A}\right\rangle _{ij}  &  =%
{\displaystyle\sum\limits_{k}}
\omega_{k}^{A}\left\vert \left\langle \psi_{ij}\right.  \left\vert \varphi
_{k}\right\rangle \right\vert ^{2}\nonumber\\
& \nonumber\\
\left\langle \hat{O}_{B}\right\rangle _{ij}  &  =%
{\displaystyle\sum\limits_{k}}
\omega_{k}^{B}\left\vert \left\langle \psi_{ij}\right.  \left\vert \varphi
_{k}\right\rangle \right\vert ^{2}%
\end{align}

\item $\hat{O}_{A}$ and $\hat{O}_{B}$ are the formal Hermitian outcome
operators%
\begin{align}
\left\langle \hat{O}_{A}\right\rangle  &  =%
{\displaystyle\sum\limits_{k}}
\omega_{k}^{A}\left\vert \varphi_{k}\right\rangle \left\langle \varphi
_{k}\right\vert \nonumber\\
& \nonumber\\
\left\langle \hat{O}_{B}\right\rangle  &  =%
{\displaystyle\sum\limits_{k}}
\omega_{k}^{B}\left\vert \varphi_{k}\right\rangle \left\langle \varphi
_{k}\right\vert
\end{align}

\item Ranking the expected outcomes in numerical order gives each an induced
preference on the possible pre-measurement output states. It is these induced
preferences which the players analyse in order to determine their choice of
action from the sets $\left\{  \hat{\alpha}_{1},\ldots,\hat{\alpha}%
_{m}\right\}  $ and $\left\{  \hat{\beta}_{1},\ldots,\hat{\beta}_{n}\right\}
$ and thus the induced preferences determine the game the players
\textit{actually play}, even though the weightings assigned to the
\textit{measurement results} may reflect a different game
\end{itemize}

This structure defines induced preferences based on a calculation of the
expected outcomes. One could, for example, also envision inducing preferences
on the pre-measurement output states by the calculation of some expectation of
a function of the output operators $\left\langle f\left(  \hat{O}_{A}\right)
\right\rangle $ and $\left\langle f\left(  \hat{O}_{B}\right)  \right\rangle
$. As we have noted above, the players are free to determine their preferences
in whatever way they see fit. Once their preferences have been fixed, and
assuming rational players, it is then a computational matter to determine what
their best choice of action is (if one exists). In games of complete
information each player can thus predict the action of the other where a fixed
(equilibrium) action exists.

\subsection{Preferences Based on a Distance Measure}

A measure that yields a geometric interpretation is that of distance. Thus we
might suppose that player $A$ wishes to get as close as possible, as
determined by this distance measure, to some state $\left\vert \pi
_{A}\right\rangle $ whereas player $B$ is trying to minimize the distance of
the output states to some other state $\left\vert \pi_{B}\right\rangle $. Of
course, it may be that $\left\vert \pi_{A\left(  B\right)  }\right\rangle
\notin\Psi_{out}$, so the players rank the possible output states in terms of
increasing $\left\vert \left\langle \psi_{ij}\right.  \left\vert \pi_{A\left(
B\right)  }\right\rangle \right\vert ^{2}$ if the square overlap is used as
the distance measure.

Let us consider the set $\Psi$ of all states for the physical system upon
which the players act. For a DQG we clearly have $\Psi_{out}\subset\Psi$. That
is, the actions available to the players do not produce a map of the input
state onto the complete set of possible states of the physical system. We now
consider 3 ways of defining preferences on the set of possible pre-measurement
output states:

\begin{enumerate}
\item \textbf{P1} : We define a preference \textit{directly} on the output
states by ordering them according to their distance from some \textit{global}
most preferred state. The global maximum is the state we would most prefer to
be the output given a choice over the \textit{entire} Hilbert space of outputs
$\Psi$. This global most preferred state will be the state that maximises the
expected outcome $\left\langle \hat{O}_{A\left(  B\right)  }\right\rangle $
for \textit{some} measurement and associated numerical weights.

\item \textbf{P2} : We define a preference \textit{directly} on the output
states by ordering them according to their distance from some \textit{local}
most preferred state, where this is defined as the output state from the set
of possible output states that minimizes the distance to the globally most
preferred state defined in preference \textbf{P1}. Thus the distance is
defined strictly in terms of the possible output states which are all
contained within $\Psi_{out}$

\item \textbf{P3} : We define a set of preferences over the measurement
eigenstates of some operator, as expressed by some weightings. These then
\textit{induce} a preference over the output states by calculation of the
expected outcomes for each possible output state in $\Psi_{out}$. Note that,
if we denote the set of measurement eigenstates by $\Psi_{meas}$ then it is
possible to have $\Psi_{meas}\cap\Psi_{out}=\emptyset\bigskip$
\end{enumerate}

We denote the \textit{global} state (the state in the universal set $\Psi$)
that maximises $\left\langle \hat{O}_{A\left(  B\right)  }\right\rangle $ by
$\left\vert \Gamma_{A\left(  B\right)  }\right\rangle $ and the \textit{local}
state (the state in the restricted set $\Psi_{out}$) that maximises
$\left\langle \hat{O}_{A\left(  B\right)  }\right\rangle $ by $\left\vert
\gamma_{A\left(  B\right)  }\right\rangle $\footnote{Note that $\hat
{O}_{A\left(  B\right)  }$ are not specified; we merely note that such
operators representing measurements exist. Indeed, one of the challenges when
we work with preferences on pre-measurement states is to construct the
observables and weightings that realize those preferences.}. Because the
available operations of the players do not map the input state to $\Psi$ the
local maximum state does not necessarily equal the global maximum state. If
the \textit{actual} state produced by the players is $\left\vert \psi
_{out}\right\rangle \in\Psi_{out}$ then we have%
\begin{align}
\left\langle \hat{O}_{A}\right\rangle  &  =%
{\displaystyle\sum\limits_{k}}
\omega_{k}^{A}\left\vert \left\langle \psi_{out}\right.  \left\vert
\varphi_{k}\right\rangle \right\vert ^{2}\leq%
{\displaystyle\sum\limits_{k}}
\omega_{k}^{A}\left\vert \left\langle \gamma_{A}\right.  \left\vert
\varphi_{k}\right\rangle \right\vert ^{2}\leq%
{\displaystyle\sum\limits_{k}}
\omega_{k}^{A}\left\vert \left\langle \Gamma_{A}\right.  \left\vert
\varphi_{k}\right\rangle \right\vert ^{2}\nonumber\\
& \nonumber\\
\left\langle \hat{O}_{B}\right\rangle  &  =%
{\displaystyle\sum\limits_{k}}
\omega_{k}^{B}\left\vert \left\langle \psi_{out}\right.  \left\vert
\varphi_{k}\right\rangle \right\vert ^{2}\leq%
{\displaystyle\sum\limits_{k}}
\omega_{k}^{B}\left\vert \left\langle \gamma_{B}\right.  \left\vert
\varphi_{k}\right\rangle \right\vert ^{2}\leq%
{\displaystyle\sum\limits_{k}}
\omega_{k}^{B}\left\vert \left\langle \Gamma_{B}\right.  \left\vert
\varphi_{k}\right\rangle \right\vert ^{2}\nonumber\\
& \nonumber\\
&
\end{align}
The preference definitions \textbf{P1} and \textbf{P2} are not equivalent.
This can easily be seen by consideration of the following output states
$\Psi_{out}=\left\{  \left\vert \psi_{1}\right\rangle ,\left\vert \psi
_{2}\right\rangle ,\left\vert \psi_{3}\right\rangle \right\}  $ of a game
played with a single qubit such that%
\begin{align}
\left\vert \psi_{1}\right\rangle  &  =a\left\vert 0\right\rangle +b\left\vert
1\right\rangle \nonumber\\
\left\vert \psi_{2}\right\rangle  &  =b\left\vert 0\right\rangle -a\left\vert
1\right\rangle \nonumber\\
\left\vert \psi_{3}\right\rangle  &  =c\left\vert 0\right\rangle +d\left\vert
1\right\rangle
\end{align}
with $a,b,c,d\in%
\mathbb{R}
$ such that $a^{2}>b^{2}>c^{2}$ and we assume the global most preferred state
is $\left\vert 0\right\rangle $. Defining a preference based on distance to
this \textit{globally} most preferred state leads to the preference ordering
on the outputs%
\begin{equation}
\mathbf{P1:~~~}\left\vert \psi_{1}\right\rangle \succ\left\vert \psi
_{2}\right\rangle \succ\left\vert \psi_{3}\right\rangle
\end{equation}
However, using the preference definition \textbf{P2} and defining a preference
based on the distance to the \textit{locally} most preferred state $\left\vert
\psi_{1}\right\rangle =a\left\vert 0\right\rangle +b\left\vert 1\right\rangle
$ yields the ordering%
\begin{equation}
\mathbf{P2:~~~}\left\vert \psi_{1}\right\rangle \succ\left\vert \psi
_{3}\right\rangle \succ\left\vert \psi_{2}\right\rangle
\end{equation}

Let us now consider a preference measure defined by \textbf{P3} such that we
have the preferences of a strictly competitive game defined on the
\textit{measurement eigenstates}. With a suitable labelling of these
eigenstates we can, using numerical weightings, express the preferences of
player \textit{A} as%
\begin{align}
\left\vert \varphi_{1}\right\rangle  &  \longrightarrow a\nonumber\\
\left\vert \varphi_{j}\right\rangle  &  \longrightarrow b~~~~\left(
j\neq1\right) \nonumber\\
a  &  >b
\end{align}
giving the ordering on the measurement eigenstates as%
\begin{equation}
\left\vert \varphi_{1}\right\rangle ~\succ~\left\vert \varphi_{2}\right\rangle
~=~\left\vert \varphi_{3}\right\rangle ~=~\left\vert \varphi_{4}\right\rangle
~=\ldots
\end{equation}
If we assume a ranking according to expected payoff, this preference on the
(post-measurement) eigenstates induces a preference on the (pre-measurement)
output states. The expected outcome for each possible output state is%
\begin{align}
\left\langle \hat{O}_{A}\right\rangle _{ij}  &  =%
{\displaystyle\sum\limits_{k}}
\omega_{k}^{A}\left\vert \left\langle \psi_{ij}\right.  \left\vert \varphi
_{k}\right\rangle \right\vert ^{2}~=~a\left\vert \left\langle \psi
_{ij}\right.  \left\vert \varphi_{1}\right\rangle \right\vert ^{2}+b%
{\displaystyle\sum\limits_{k=2}}
\left\vert \left\langle \psi_{ij}\right.  \left\vert \varphi_{k}\right\rangle
\right\vert ^{2}\nonumber\\
& \nonumber\\
&  =\left(  a-b\right)  \left\vert \left\langle \psi_{ij}\right.  \left\vert
\varphi_{1}\right\rangle \right\vert ^{2}+b
\end{align}
But this gives an ordering on the pre-measurement output states according to
their the distance from the most preferred post-measurement state $\left\vert
\varphi_{1}\right\rangle $. Thus the preferences on the pre-measurement states
induced by assuming strictly competitive preferences on the post-measurement
states, and a ranking according to the expected payoff, are the \textit{same}
preferences on the possible outputs that we obtain from \textbf{P1} when
$\left\vert \varphi_{1}\right\rangle \notin\Psi_{out}$ (or \textbf{P2} when
$\left\vert \varphi_{1}\right\rangle \in\Psi_{out}$). In other words, a
distance ordering on the pre-measurement output states is equivalent to
assuming the preferences of a strictly competitive game on the
post-measurement states. We thus obtain the result that if the (globally) most
preferred state of a player is an eigenstate of some Hermitian operator:

\begin{quote}
the preference defined by the distance to some state $\left\vert
\xi\right\rangle $ is equivalent to a strictly-competitive ordering, by
weighting, on the eigenstates of some Hermitian operator $\hat{M}$ such that
$\left\vert \xi\right\rangle $ is an eigenstate of $\hat{M}$ that is assigned
the biggest weight, where the expected outcome is the quantity the players
wish to maximise
\end{quote}

\subsubsection{A simple strictly-competitive game}

In order to illustrate the equivalence of distance-ordering on the outputs and
a strictly-competitive ordering on the measurement results we will consider
the 2-qubit example discussed above. In this case we have the set of output
states and the set of measurement eigenstates given by%
\begin{align}
\Psi_{out}  &  =\left\{  \left\vert \psi_{1}\right\rangle ,\left\vert \psi
_{2}\right\rangle ,\left\vert \psi_{3}\right\rangle ,\left\vert \psi
_{4}\right\rangle \right\} \nonumber\\
\Psi_{meas}  &  =\left\{  \left\vert \varphi_{1}\right\rangle ,\left\vert
\varphi_{2}\right\rangle ,\left\vert \varphi_{3}\right\rangle ,\left\vert
\varphi_{4}\right\rangle \right\}
\end{align}
As above we will assume that $\left\vert \varphi_{1}\right\rangle $ is the
most preferred measurement state and that $\left\vert \psi_{1}\right\rangle $
will denote the most preferred output state (we can always relabel the states
so that this is true). Now let us assume an ordering according to \textbf{P1}.
Let us assume that we have%

\begin{equation}
\left\vert \left\langle \Gamma_{A}\right.  \left\vert \psi_{1}\right\rangle
\right\vert ^{2}>\left\vert \left\langle \Gamma_{A}\right.  \left\vert
\psi_{2}\right\rangle \right\vert ^{2}>\left\vert \left\langle \Gamma
_{A}\right.  \left\vert \psi_{3}\right\rangle \right\vert ^{2}>\left\vert
\left\langle \Gamma_{A}\right.  \left\vert \psi_{4}\right\rangle \right\vert
^{2}%
\end{equation}
where $\left\vert \Gamma_{A}\right\rangle $ is the global maximum state, or
the state $\left\vert \Gamma_{A}\right\rangle \in\Psi$ that is the most
preferred state out of all possible states on the Hilbert space. This is
equivalent to an ordering of the output states given by%
\[
\left\vert \psi_{1}\right\rangle ~\succ~\left\vert \psi_{2}\right\rangle
~\succ~\left\vert \psi_{3}\right\rangle ~\succ~\left\vert \psi_{4}%
\right\rangle
\]
Now let us consider the preferences defined by P3. In this case we have, as
above, that
\begin{align}
\left\vert \varphi_{1}\right\rangle  &  \longrightarrow a\nonumber\\
\left\{  \left\vert \varphi_{2}\right\rangle ,\left\vert \varphi
_{3}\right\rangle ,\left\vert \varphi_{4}\right\rangle \right\}   &
\longrightarrow b\nonumber\\
a  &  >b
\end{align}
so that the preferences are, \textit{initially}, defined on the measurement
set $\Psi_{meas}$. The expected payoffs are given by%
\begin{equation}
\left\langle \hat{O}_{A}\right\rangle _{j}=a\left\vert \left\langle \psi
_{j}\right.  \left\vert \varphi_{1}\right\rangle \right\vert ^{2}+b\left(
1-\left\vert \left\langle \psi_{j}\right.  \left\vert \varphi_{1}\right\rangle
\right\vert ^{2}\right)  ~~~~~j=1,2,3,4
\end{equation}
However, in the perspective of \textbf{P3}, the global maximum state
$\left\vert \Gamma_{A}\right\rangle $ is just the most preferred state from
$\Psi_{meas}$. Thus we have that $\left\vert \Gamma_{A}\right\rangle
=\left\vert \varphi_{1}\right\rangle $. Further, by assumption we have that%

\begin{equation}
\left\vert \left\langle \Gamma_{A}\right.  \left\vert \psi_{1}\right\rangle
\right\vert ^{2}>\left\vert \left\langle \Gamma_{A}\right.  \left\vert
\psi_{2}\right\rangle \right\vert ^{2}>\left\vert \left\langle \Gamma
_{A}\right.  \left\vert \psi_{3}\right\rangle \right\vert ^{2}>\left\vert
\left\langle \Gamma_{A}\right.  \left\vert \psi_{4}\right\rangle \right\vert
^{2}%
\end{equation}
which is equivalent to the ordering%

\begin{equation}
\left\vert \left\langle \varphi_{1}\right.  \left\vert \psi_{1}\right\rangle
\right\vert ^{2}>\left\vert \left\langle \varphi_{1}\right.  \left\vert
\psi_{2}\right\rangle \right\vert ^{2}>\left\vert \left\langle \varphi
_{1}\right.  \left\vert \psi_{3}\right\rangle \right\vert ^{2}>\left\vert
\left\langle \varphi_{1}\right.  \left\vert \psi_{4}\right\rangle \right\vert
^{2}%
\end{equation}
which gives the ordering of the \textit{expected} payoffs%

\begin{equation}
\left\langle \hat{O}_{A}\right\rangle _{1}>\left\langle \hat{O}_{A}%
\right\rangle _{2}>\left\langle \hat{O}_{A}\right\rangle _{3}>\left\langle
\hat{O}_{A}\right\rangle _{4}%
\end{equation}
which is, of course, equivalent to an ordering on the output states of
\begin{equation}
\left\vert \psi_{1}\right\rangle ~\succ~\left\vert \psi_{2}\right\rangle
~\succ~\left\vert \psi_{3}\right\rangle ~\succ~\left\vert \psi_{4}%
\right\rangle
\end{equation}
Thus, for this strictly-competitive example, the preference definitions
\textbf{P1} and \textbf{P3} are entirely equivalent; in both cases we obtain
an ordering \textit{of the output states} that is dependent of the distance of
the output states from the most preferred global state, where distance is
expressed by the projection $\left\vert \left\langle \psi_{j}\right.
\left\vert \varphi_{1}\right\rangle \right\vert ^{2}$. Note that in
\textbf{P1} this preference is defined directly on the output states whereas
in \textbf{P3} this preference is \textit{induced} on the output states. Thus,
for this example, defining the preferences by a preference over the
\textit{measurement} results as%

\begin{equation}
\left\vert \varphi_{1}\right\rangle ~\succ~\left\vert \varphi_{2}\right\rangle
~=~\left\vert \varphi_{3}\right\rangle ~=~\left\vert \varphi_{4}\right\rangle
\end{equation}
yields an ordering of the \textit{output states} that is the same ordering
obtained from \textbf{P1}.

\subsubsection{Preferences Based on Variances}

Here we suppose that player $A$ assigns his preference on the
(pre-measurement) output states by requiring that the variance of these output
states with respect to some subsequent measurement $\hat{M}_{A}$ is minimized.
Player $B$ seeks to produce an output state that minimizes the variance with
respect to some measurement $\hat{M}_{B}$. It is important to emphasize that
the players must have some foundation upon which to base their theoretical
analysis leading to their choice of `move' in the game. Clearly this
theoretical analysis is undertaken \textit{before} any measurement is actually
performed. Equally clearly, any physical instantiation of the game (that is,
to actually play the game with physical objects and processes) will require
\textit{some} measurement to be performed. Here we are supposing that the
foundation is supplied by a preference for minimizing variances of the
pre-measurement ouput states with respect to specific observables.

For convenience we shall suppose that $\hat{M}_{A}$ and $\hat{M}_{B}$ are
complementary observables. If the eigenstates of these operators are
$\left\vert \varphi_{i}^{A}\right\rangle $ and $\left\vert \varphi_{j}%
^{B}\right\rangle $, respectively, then the eigenstates are related by%
\begin{equation}
\left\vert \varphi_{i}^{A}\right\rangle =\frac{1}{\sqrt{N}}\sum\limits_{j=1}%
^{N}\exp\left(  i\theta_{j}\right)  \left\vert \varphi_{j}^{B}\right\rangle
\end{equation}
Player $A$ is thus wishing to produce an output state in a game that has the
smallest distance to an eigenstate of $\hat{M}_{A}$. Equivalently player $A$
wishes to maximise the distance to eigenstates of $\hat{M}_{B}$. The players
thus generate opposite preferences based on distance to some preferred set of
states. We can realize these preferences by setting up a strictly-competitive
game as we have discussed above.

\section{Discussion}

In this paper we have considered a new perspective on quantum games by
considering preferences on pre-measurement output states produced by the
players. Once the operations (`moves') available to the players are fixed
then, for a given input state, these pre-measurement preferences determine the
game that is played. In typical formulations of quantum games where weightings
are assigned to measurement results we have demonstrated that these
\textit{induce} preferences on the pre-measurement output states [4]. It is
these \textit{induced} preferences the players use to determine their choice
of strategy. In this sense, then, consideration of pre-measurement preferences
is more fundamental since these determine the actual game that the players play.

We have described a formalism for a discrete quantum game defined by the
pre-measurement preferences (for a given input state and set of available
operations). Knowledge of these is sufficient to determine the choice of
strategy and thus specifies the game. For a 2-player DQG in which there are
$mn$ possible pre-measurement output states the number of possible games that
can be played for a given input state and set of available operations is
$\sim\left(  mn!\right)  ^{2}$ since a player's preference is simply a
particular ordering of the set $\Psi_{out}$. There are choices of ordering
(preference) which are more physically appealing than others and we have
discussed examples of these. If we wish to actually play a game then some
measurement must be performed. In general then, the challenge is to find a
measurement and weightings of the measurement results which will induce the
pre-measurement preferences. We have shown that for 2 important and physically
relevant choices of preferences on the pre-measurement output states, a
minimization of distance and variance, the playable game can be realized by a
strictly competitive weighting on the measurement results.

If the players wish to minimize the distance to the states $\left\vert \pi
_{A}\right\rangle $ and $\left\vert \pi_{B}\right\rangle $, respectively, then
this is equivalent to player $A$ wishing to maximise
\textit{distinguishability} of the the output state of the game from the state
$\left\vert \pi_{B}\right\rangle $ and vice versa for player $B$. Playing such
a game would then, in an appropriate sense, produce an equilibrium state
(where one exists) that optimises the distinguishability from both $\left\vert
\pi_{A}\right\rangle $ and $\left\vert \pi_{B}\right\rangle $ for the given
set of output states that is produced.

In our previous work we have considered a distance based measure on output
states and shown its importance in strictly competitive games. Here we have
shown that a distance based preference measure on pre-measurement output
states is \textit{equivalent} to a strictly-competitive game defined on the
measurement results. If a particular quantum computation can be identified as
the equilibrium state of a game with a distance measure then we have a simple
methodology for physically constructing such a computation as a strictly
competitive game on measurement results.

\section{References}

\begin{enumerate}
\item D. Meyer, Quantum Strategies, \textit{Physical Review Letters},
\textbf{82}, pages 1052-1055.$\left(  1999\right)  $

\item J. Eisert, M. Wilkens, and M. Lewenstein, Quantum Games and Quantum
Strategies, \textit{Physical Review Letters}, \textbf{83}, pages 3077-3080
$\left(  1999\right)  $.

\item F.S. Khan and S.J.D .Phoenix, Gaming the Quantum, \textit{Journal of
Quantum Information and Computing}, \textbf{13(3-4)}, 231-244 (2013)

\item S.J.D .Phoenix and F.S. Khan, The Role of Correlation in Quantum and
Classical Games, \textit{Fluctuation and Noise Letters}, \textbf{12}, 1350011 (2013)

\item F.S. Khan and S.J.D .Phoenix, Mini-maximizing Two Qubit Quantum
Computations, \textit{Journal of Quantum Information and Computing},
\textbf{12(12)}, 3807-3819 (2013)

\item R. Cleve, P. Hoyer, B. Toner and J. Watrous, Consequences and Limits of
Nonlocal Strategies, preprint available at
http://arxiv.org/abs/quant-ph/0404076v2 (2004)

\item J. Shimamura, S. K. \"{O}zdemir, F. Morikoshi and N. Imoto, Entangled
states that cannot reproduce original classical games in their quantum
version, \textit{Physics Letters A}, \textbf{328}, 20-25 (2004)

\item A. Iqbal and S. Weigert, Quantum correlation games, \textit{Journal of
Physics A}: Math. \& Gen. \textbf{37}, 5873-5885 (2004).

\item A.P. Flitney, D. Abbott D, Quantum games with decoherence,
\textit{Journal of Physics A}: Mathematical and General \textbf{38}: 449 (2005)

\item C. F. Lee and N. Johnson, Efficiency and Formalism of Quantum Games,
\textit{Physical. Review A}, \textbf{67}, 022311 (2003)

\item T. Cheon and I. Tsutsui, Classical and Quantum Contents of Solvable Game
Theory on Hilbert Space, \textit{Physics Letters A} \textbf{348} 147-152 (2006)

\item S. E. Landsburg, Nash Equilibria in Quantum Games, \textit{Proceedings
of the American Mathematical Society,} \textbf{139}, 4423-4434, (2011)

\item S. A. Bleiler, A Formalism for Quantum Games I - Quantizing Mixtures,
preprint at http://arxiv.org/abs/0808.1389, 2008.
\end{enumerate}

\end{document}